# The Nature of Electron Transport and visible light absorption in Strontium Niobate - A Plasmonic Water Splitter

D. Y. Wan[1,2], Y. L. Zhao[1]*, Y. Cai[7], T. C. Asmara[3], Z. Huang[1], J. Q. Chen[1], J. Hong[8], C. T. Nelson[9], M. R. Motapothula[1], B. X. Yan[1,2], R. Xu[8], H. Zheng[9,10], Ariando[1,2], A. Rusydi[2,3], A. M. Minor[9,10], M. B. H. Breese[2,3], M. Sherburne[6], M. Asta[10], Q-H. Xu[1,5], T. Venkatesan[1,2,4,6,7]*

[1]NUSNNI-NanoCore, National University of Singapore, 117411, Singapore
[2]Department of Physics, National University of Singapore, 117542, Singapore
[3]Singapore Synchrotron Light Source, National University of Singapore, Singapore 117603, Singapore.
[4]Department of Electrical and Computer Engineering, National University of Singapore, 117576, Singapore
[5]Department of Chemistry, National University of Singapore, 117543, Singapore
[6]Department of Material Science and Engineering, National University of Singapore, 117575, Singapore
[7]NUS Graduate School for Integrative Sciences & Engineering, National University of Singapore, 117456, Singapore
[8]School of Chemical & Biomedical Engineering, Nanyang Technological University, 637459 Singapore
[9]Materials Science Division, Lawrence Berkeley National Laboratory, Berkeley, California, 94720, United States
[10]Department of Materials Science and Engineering, University of California, Berkeley, California, 94720, United States

*Corresponding author. Email: Venky@nus.edu.sg; yongliangrosy@gmail.com.

**Abstract**

Semiconductor compounds are widely used for water splitting applications, where photo-generated electron-hole pairs are exploited to induce catalysis. Recently, powders of a metallic oxide ($Sr_{1-x}NbO_3$, $0.03 < x < 0.20$) have shown competitive photocatalytic efficiency, opening up the material space available for finding optimizing performance in water-splitting applications. The origin of the visible light absorption in these powders was reported to be due to an interband transition and the charge carrier separation was proposed to be due to the high carrier mobility of this material. In the current work we have prepared epitaxial thin films of $Sr_{0.94}NbO_{3+\delta}$ and found that the bandgap of this material is ~4.1 eV, which is very large. Surprisingly the carrier density of the conducting phase reaches $10^{22}$ cm$^{-3}$, which is only one order smaller than that of elemental metals and the carrier mobility is only 2.47 cm$^2$/(V·s). Contrary to earlier reports, the visible light absorption at 1.8 eV (~688 nm) is due to the bulk plasmon resonance, arising from the large carrier density, instead of an interband transition. Excitation of the plasmonic resonance results in a multifold enhancement of the lifetime of charge carriers. Thus we propose that the hot charge carriers generated from decay of plasmons produced by optical absorption is responsible for the water splitting efficiency of this material under visible light irradiation.

**Introduction**

Converting solar energy into chemical energy (e.g. splitting water by sun light) with the aid of photocatalysts is a promising way to reduce the increasing demand for fossil fuels[1, 2, 3, 4, 5, 6, 7, 8, 9, 10]. Very few oxide semiconductors have been used as photocatalysts since they need to be chemically robust and their bandgap should be neither too wide nor narrow in order to absorb sun light in the visible range efficiently and also satisfy the minimum energy requirement (1.23 eV theoretically but >1.9 eV experimentally) for splitting water into hydrogen and oxygen[11, 12, 13]. Large bandgap oxides (such as $TiO_2$) are used in photocatalytic water splitting either by reducing their optical bandgap to absorb visible light or incorporating visible light absorbers such as organic dyes, low bandgap quantum absorbers or metal nanostructures[14, 15] in the host. In the former case, cationic or anionic doping or combination of both are typically applied to narrow the bandgap[16, 17, 18, 19]. Unfortunately, in most cases the bandgap change achieved is small mainly due to the fact that the discrete defect states are normally very close to the band edges[20, 21, 22]. In the latter case, hot electrons in the visible light absorber inject into the conduction band of large bandgap oxides, which are subsequently used to reduce water to hydrogen gas. Among the visible light absorbers, metal (e.g. Au, Ag) nanostructures are special as the hot electrons generated by the decay of visible light excitation of surface plasmon resonance (SPR) can be injected in to a large bandgap semiconductor like $TiO_2$[23, 24]. However, the materials used for enhancing the photocatalytic activity by using SPR are mainly Au, Ag nanostructures, which are not low cost materials.

Recently a red metallic oxide $Sr_{1-x}NbO_3$ (0.03 < x < 0.20) (in the form of powders) was used in photocatalytic water splitting and the authors proposed a special band structure, in which the electron-hole pairs come from the optical transition from the metallic conduction band (about 1.9 eV above the valence band) to a higher level unoccupied band[25, 26]. In this report, the visible light absorption was attributed to the electron's interband transitions and the electron-hole pair separation was attributed to the assumption of high carrier mobility although only temperature dependent conductivity was measured. As both charge carrier density and mobility contribute to the conductivity of a sample, simply assuming a large mobility for a highly conductive material may lead to a wrong conclusion. Hence obtaining the mobility as well as other electrical transport properties is crucial for understanding the details of the photo-generated carrier separation process. The optical bandgap obtained from Kubelka-Munk transformation of the reflectance spectrum of the powder form $Sr_{1-x}NbO_3$ is inaccurate since it neglects the plasmonic absorption. The proper way to measure the optical bandgap is to obtain the complex dielectric function, by using Kramers-Kronig-transformed reflectivity or spectroscopic ellipsometry. Epitaxial thin films are required for measuring the intrinsic mobility since the grain boundaries in the film are much less than in powders. Using thin films also allows us to measure the transmittance, reflectance and ellipsometry spectra accurately, which can give proper optical and plasmonic absorptions of this material. Furthermore, both the band structure and the process of the hot electron transition of this material can also be investigated by femtosecond time resolved transient absorption (TA) spectroscopy.

Here we have prepared $Sr_{0.94}NbO_{3+\delta}$ films by pulsed laser deposition (PLD) at various oxygen partial pressures on top of insulating $LaAlO_3$ substrates and compared their optical spectra, electronic transport and carrier dynamics properties. We found the optical bandgap to be around 4.1 eV, which was almost independent of the oxygen content although the crystal structure changed from pseudo-tetragonal perovskite to orthorhombic with increasing oxygen partial pressure from 5×10$^{-6}$ Torr to 1×10$^{-4}$ Torr. The bulk plasmon peak for the film grown at 5×10$^{-6}$ Torr was found at about 1.8 eV (688 nm), which is at the appropriate energy for solar-photocatalytic water splitting. The high conductivity (~10$^4$ S/cm) of the sample prepared at low pressure is mainly contributed by the high charge carrier density (~10$^{22}$ cm$^{-3}$) rather than its mobility (2.47 cm$^2$/ (V·s)) at room temperature. Thus we believe that the photocatalytic activity of Sr$_{0.94}$NbO$_{3+\delta}$ under visible to near-infrared irradiation is due to the hot electrons generated from the decay of the plasmon in Sr$_{0.94}$NbO$_3$ instead of interband absorption transition. Thus Sr$_{0.94}$NbO$_{3+\delta}$ represents an extraordinary material system, which has a large bandgap of 4.1 eV but a degenerate conduction band with a large carrier density exceeding 10$^{22}$ electrons/ cm$^3$ which leads to strong useful plasmonic effects.

**Results and discussions**

The XRD spectrum of the film deposited at oxygen partial pressure of 5 × 10$^{-6}$ Torr is shown in Fig. 1. The θ-2θ scan indicates the film's lattice parameter along out-of-plane [001] direction as 4.10 Å. The full width at half maximum (FWHM) of the rocking curve is measured to be 0.71 Å, which is acceptable by considering the large lattice mismatch between the film and the substrate (LaAlO$_3$ forms in pseudocubic perovskite structure at room temperature with lattice constant 3.79 Å). The reciprocal space maps (RSMs) of (-103) and (103) [(0-13) and (013)] are symmetric with respect to the out-of-plane [001] axis, suggesting orthogonality of [001] and [100] ([001] and [010]) axes. Despite their relatively broad RSM spots, the in-plane parameters are obtained as 4.04 Å equally. Hence the film forms in tetragonal-like structure on LaAlO$_3$ substrate with a large strain near the interface. The strain effect can be clearly seen in the local HRTEM image of the lattice (Fig. 1(c)). The highlighted open burgess circuit indicates an a[100]$_p$ type edge dislocation core, where the extra plane on the LaAlO$_3$ side indicates a compressive misfit of the film. As the oxygen partial pressure increases to 1 × 10$^{-4}$ Torr, a small shift of the film peaks towards low angles can be observed in the θ-2θ scan and the FWHM of the rocking curve increases (Fig. S1). Structural changes from tetragonal to orthorhombic were observed in the local HRTEM images (Fig. S2). The orthorhombic structure is close to the reported structure of Sr$_2$Nb$_2$O$_7$ (bulk a=3.933 Å, b=26.726 Å, c=5.683 Å), which has an equivalent tetragonal structure with lattice parameters a=b=3.901 Å and c=3.933 Å. The decreasing of the out-of-plane and in-plane lattice parameters with oxygen partial pressure was obtained from the electron diffraction pattern (Fig. S2, Table S1). At the intermediate oxygen partial pressure, mixed structure exists in the film.

It was reported that Sr content strongly determines the crystal structure of the non-stoichiometric SrNbO$_3$ phase[27, 28]. Here the elemental content of the films deposited at different oxygen partial pressures are precisely studied (Fig. S3). The cationic contents are identical for the films within the detection limit of PIXE and the Sr/Nb atomic ratio is measured as 0.94/1. The deficiency of Sr content comes from the variation in the target preparation process[29].

A cut off of the transmission edge is observed near 300 nm (Fig. 2(a)), which indicates an optical bandgap of ~ 4.1 eV (from Tauc plot-indirect, Fig. S4) and it is almost independent of the preparation oxygen pressure. Both the transmission and reflection of the film prepared at 5 × 10$^{-6}$ Torr were plotted, from which the accurate absorption spectrum could be obtained (Fig. 2(b)). The minimum reflection is located at around 600 nm, which could indicate the rough frequency of its volume plasmon. Hence the reflection spectrum between 500 nm and 1000 nm can be well fit by the Drude Model, and the corresponding plasmon frequency of the fitting curve is 1.6 eV, which has a small difference with the plasmon frequency measured by spectroscopic ellipsometry. The transmission of the films continuously increases with oxygen partial pressure above 600 nm, which indicates absorption along with free carrier absorption (Drude Model) in this wavelength range, where the latter is consistent with the metallic nature of the films and powders[25, 30].

The complex refractive index, $\tilde{n}(\omega) = n(\omega) + i\kappa(\omega)$, and the loss function, -Im[$\varepsilon^{-1}(\omega)$], spectra of the 5x10$^{-6}$ Torr sample extracted from spectroscopic ellipsometry data are shown in Fig. 2(c) and 2(d). The extinction coefficient spectrum, $\kappa(\omega)$, of the sample (Fig. 2(d)) shows that it has a Drude peak below 2 eV (typical of a metal) and a first interband transition peak (indicating the bandgap of the film) above 4.1 eV, consistent with its transmission spectrum (Fig. 2(b)). Between these two peaks, the $\kappa(\omega)$ is featureless, indicating the lack of major optical transitions within the 2 – 4.1 eV energy range. Meanwhile, the loss function spectrum of the sample (Fig. 2(d)) shows a large peak at ~1.5 - 2.1 eV with a peak position of ~1.8 eV (688 nm), indicating the existence of a bulk plasmon at that energy[31]. From n($\omega$) and $\kappa(\omega)$ spectra, the normal-incident reflectivity of the film can be obtained using Fresnel equations, as shown in Fig. 2(e). This reflectivity is consistent with the spectrum measured by UV-Visible spectroscopy. From the reflectivity, the Kubelka-Munk function of the film can be obtained (Fig. 2(e)), and it can be seen that the shape of the function resembles the previous reported results, with an apparent absorption edge at ~2 eV. Since there is no peak in the κ(ω) spectrum at around that energy, this absorption edge does not come from intra- or interband transition as previously reported. Instead, this absorption edge is plasmonic in origin because it coincides with the plasmon peak at ~1.8 eV in the loss function spectrum.

The origin of hot electrons in SrNbO$_3$ under irradiation with visible light, which can be used for water splitting, can be interpreted by the plasmon model. When SrNbO$_3$ is under irradiation, there will be a resonant collective oscillation of the electrons in the conduction band, which is the surface plasmon. Hot electrons are generated by the decay of the plasmon and transferred to co-catalysts (e.g. Pt), where the H$^+$ reduction reaction can take place. The holes left in the SrNbO$_3$ can drive the oxidation reaction at the surface of this material. However, it was reported that the visible light absorption was attributed to the electrons' interband transitions and the electron-hole pair separation was attributed to a possible high mobility of the electrons, which is inconsistent with our model. To resolve this problem, the carrier transport properties were studied for this material.

The conducting property of the film is strongly dependent on the oxygen partial pressure, where a transition of metallic to semiconductor transport behavior is clearly seen when increasing the deposition oxygen partial pressure (Fig. 3(a)). The electron density of the most conductive sample (prepared at 5 × 10$^{-6}$ Torr) reaches 10$^{22}$cm$^{-3}$ and it is almost independent of the measurement temperature, which agrees with the reported data[32](Fig. 3(b)) indicative of a degenerate Fermi level. In contrast, the electron mobility is only 2.47 cm$^2$/(V·s) at room temperature, which is not outstanding compared with other oxides (e.g. TiO$_2$, BaSnO$_3$) and semiconductors[33, 34, 35, 36] (Fig. 3(c)). So the high conductivity of this material is due to the high carrier density and not the carrier mobility. The absence of significant internal electric field to avoid electron-hole recombination also implies that an interband transition model is not suitable. As the oxygen content increases in the film, the sample becomes more insulating. Both the charge carrier density and the mobility decreases with oxygen partial pressure, which is consistent with the observed two crystal structures of the materials[37]. Metallic $Sr_{0.94}NbO_3$ forms in tetragonal perovskite structure at $5\times10^{-6}$ Torr and almost insulating $Sr_{0.94}NbO_{3.4}$ forms in orthorhombic structure at $1\times10^{-4}$ Torr. At the intermediate pressure, the semiconductor forms in cermet structure.

To further understand the role of the plasmon in the catalytic process, the transient absorption spectroscopy and time-resolved pump-probe spectroscopy were used to characterize the carrier dynamic process in strontium niobate. The Fig. 3 (d) and (e) show the various excitation wavelength dependent differential reflectance (ΔR/R) spectra with different delay time. Two peaks located near 600 nm (positive ΔR/R, 2.07 eV) and 670 nm (negative ΔR/R, 1.85 eV) are observed in the transient reflection spectra. The sign of differential reflection signal would be usually opposite to the sign of differential transmission signal [38, 39]. Therefore, the positive 600 nm peak can be attributed to the optical absorption of the excited electrons (hot electrons), which might be the transition from the valence band to the deep trapped states. The negative 670nm peak can be attributed to both the transition from deep trapped states to conduction band and the optical absorption of plasmonic resonance which is consistent with the plasmonic resonance peak measured by the ellipsometry spectroscopy. It should be noted that though the transient absorption spectrum can show the signal from deep trap states, the transition process related with these states cannot be used for the photocatalytic water splitting process because the deep trapped states can only act as the recombination centers for photo generated electron-hole pairs. The intensity of differential transient reflection spectra excited by 685 nm pump light at 0.5 ps delay is very strong compared with that of other delays which can be attributed to the strong absorption of the plasmon resonance.

Fig. 4 (a) shows the Landau decay process of plasmon resonance [40]. The Landau decay process is very short which is in the time scale of a few hundred femtoseconds. The lifetime of Landau process as a function of excitation wavelength is presented in Fig. 4 (c). The lifetime increases to its maximum when the excitation wavelength is 685 nm which is near the plasmon peak of $SrNbO_3$. Fig. 4 (b) shows decay curve of the transition from deep trapped states to the conduction band. However, because the density of unfilled states in the conduction band depends on the thermal dissipation process of hot carriers, thus this decay lifetime corresponds to the thermal dissipation process of hot carriers in $SrNbO_3$ film. We can see the lifetime of this process can be as long as 400 to 550 ps when it is excited by the pump pulse with energy higher than that of the plasmon, while it will decrease to about 250 ps when the pump energy is lower than that of the plasmon. This shows that plasmon resonance can increase the lifetime of hot carrier thermal dissipation process which may also enhance the photocatalytic activity of $SrNbO_3$ as the lifetime of the hot electrons is long enough for the carriers to convert water into gases before recombination.

The energy band structures of $SrNbO_3$, $SrNbO_{3.4}$ and $SrNbO_{3.5}$ are calculated using density functional theory (DFT) and shown in Fig. 5. In the calculations, the perovskite structure was assumed for the stoichiometric SrNbO$_3$ compound, and the extra oxygen atoms for the hyperstoichiometric compositions were assumed to order into planar defects, as illustrated by the structural figures in the left panels of Fig. 5. This structural model is consistent with electron microscopy analyses that will be reported elsewhere[41, 42]. The Fermi level of SrNbO$_3$ is located in the conduction band, which implies that this material is metallic even though the bandgap is as large as 4.1 eV (we note the calculations predict a smaller bandgap relative to this experimental value, as is typically found from DFT). The Fermi level of SrNbO$_{3.4}$ is located near the bottom of the conduction band, so the conductivity is poorer than that of SrNbO$_3$ as there are fewer states for the free carriers leading to a lower carrier density. These results are consistent with the experimental measurements. Unlike SrNbO$_3$ and SrNbO$_{3.4}$, the Fermi level of SrNbO$_{3.5}$ is located at the top of the valence band, so SrNbO$_{3.5}$ is insulating and the film should be transparent, consistent with experiments.

In summary, we have demonstrated that epitaxial, single crystal Sr$_{0.94}$NbO$_3$ film can be obtained by PLD. The electron's mobility of this material is very normal, only 2.47 cm$^2$/(V·s) at room temperature, so the interband transition model cannot be applied to explain the photocatalytic activity of this material. Further, the material has a degenerate conduction band with a gap of 4.1 eV but the large carrier density leads to a large bulk plasmon at 1.6-1.8 eV which simulates a mid-gap absorption. The lifetime of the plasmon in this material is very long (~200 ps) and hence we propose that the hot electrons generated from the decay of plasmons produced by absorption of sunlight, should be responsible for the photocatalytic activity of this material. Understanding of this first plasmonic metallic oxide and its use as a photocatalyst will open the doors for the design of a new family of photocatalytic materials.

**Methods**

**Material preparation:** The PLD target was prepared by solid reactions of $Sr_4Nb_2O_9$ precursor, Nb (Alfa Aesar, 99.99%, -325 meshes) and $Nb_2O_5$ (Alfa Aesar, 99.9985%, metals basis) powder mixtures in the proper molar ratio. The precursor was prepared by calcining $SrCO_3$ (Alfa Aesar, > 99.99%, metals basis) and $Nb_2O_5$ powder mixtures in a molar ratio of 4:1. The calcination and sintering were done in air and Ar gas environment for 20 hours at a temperature of 1200 °C and 1400 °C respectively. The films of $SrNbO_3$ were deposited on $LaAlO_3$ substrate from these targets by Pulsed Laser Deposition (PLD) where a Lambda Physik Excimer KrF UV laser with wavelength of 248 nm was used. The films were deposited at 750 °C, laser energy density of 2 J/cm$^2$, laser frequency of 5 Hz and oxygen partial pressure of $5 \times 10^{-6}$ Torr to $1 \times 10^{-6}$ Torr. Typically, 130 nm thick film could be obtained with half an hour deposition.

**Physical characterization:** The elemental compositions of the films were studied by 2 MeV proton induced X-ray emissions (PIXE) with Si (Li) detector and 15 MeV carbon ions Rutherford Backscattering Spectrometry (RBS). The obtained spectra were precisely fitted by SIMNRA simulation software. The crystal structure of the film was studied by high resolution X-ray diffraction (XRD, Bruker D8 with Cu Kα1 radiation, λ = 1.5406 Å) together with the reciprocal space maps (RSMs). The local structure was measured by High Resolution Scanning Transmission Electron Microscopy (HRSTEM: FEI Titan (Team0.5)@300 kV). Optical bandgap of the film was measured by UV-visible spectrophotometer (Shimadzu SolidSpec-3700). The transmissions of the films were measured and the corresponding absorption coefficients at particular wavelengths were derived from Beer-Lambert Law. Physical Properties Measurement System (PPMS, Quantum Design Inc.) was used to measure the electrical transport properties.

**Photocatalytic water splitting reaction:** Light source used in present experiment is 300 W Xe lamp with 420 nm filter. $H_2$ evolution was measured by suspending 50 mg sample powders together with 1 wt. % Pr co-catalyst in 100 ml oxalic acid aqueous solution (0.025 M). The evolved gases were collected and quantified by an online gas chromatograph (Agilent 6890N, Argon as carrier gas, 5 Å molecular sieve column, and TCD detector).

**Transient absorption measurement:** The electronic band structure of the films (all prepared at $5 \times 10^{-6}$ Torr, on $LaAlO_3$ substrate) and the lifetimes of the photo exactions were investigated by femtosecond time resolved transient absorption (TA) spectroscopy. The laser pulses were generated using a mode-locked Ti:sapphire oscillator seeded regenerative amplifier with a pulse energy of 2 mJ at 800 nm and a repetition rate of 1 kHz. The laser beam was split into two portions. The larger portion of the beam passed through a Light Conversion TOPAS-C optical parametric amplifier to generate 350 nm as the pump beam. The intensity of the pump beam was attenuated using a neutral density filter and modulated using an optical chopper at a frequency of 500 Hz. The smaller portion of the beam was used to generate white light by passing through a 1 mm sapphire plate, which acted as the probe beam. The white light beam was further split into two portions: one was used as the probe and the other was used as the reference to correct for the pulse-to-pulse intensity fluctuation. The pump beam was focused onto the sample surface with a beam size of 300 µm, and it fully covered the smaller probe beam (diameter: 100 µm). The reflection of the probe beam from the sample surface was collected with a pair of lens and focused into a spectrometer. Very thick film samples were used to minimize the signal contribution from the substrate. The delay between the pump and the probe pulses was controlled by a computer-controlled translation stage (Newport, ESP 300). Pump probe experiments were carried out at room temperature. During the measurements, the pump and the probe energies were kept low enough to minimize damage to the samples.

**Theoretical calculations:** The atomic and electronic structure of $SrNbO_{3+x}$ compounds were performed employing spin-polarized density-functional-theory (DFT) calculations, using the Perdew-Burke-Ernzerhof (PBE96)[43] exchange-correlation potential, and the projector-augment wave (PAW) method[44, 45], as implemented in the Vienna ab-initio simulation program (VASP)[46]. In these calculations Sr $4s4p5s$, Nb $4p5s4d$, and O $2s2p$ orbitals were treated as valence states, employing the PAW potentials labeled "Sr_sv", "Nb_pv" and "O" in the VASP PBE library. The cutoff energy for the plane-wave basis set was set to 450 eV, and the DFT+U approach due to Dudarev et al.[47] was employed to treat the Nb $4d$ orbitals occupied in the $Nb^{4+}$ ions present for $SrNbO_3$ and $SrNbO_{3.4}$ compositions with the value of $U - J$ set to 4 eV. $SrNbO_3$, $SrNbO_{3.4}$, $SrNbO_{3.5}$ were modeled by supercells containing, respectively, 20 atoms with space group Pnam, 54 atoms with space group Pnnm, and 44 atoms with space group Cmc2. For the cells with oxygen excess (i.e., $SrNbO_{3.4}$ and $SrNbO_{3.5}$) the extra oxygen ions were placed in planar defects as illustrated in Fig. 5. In the structural relaxations we employed using 8*8*4, 1*4*6 and 1*4*6 k-point meshes, for $SrNbO_3$, $SrNbO_{3.4}$, $SrNbO_{3.5}$, respectively. The density of states was calculated with 16*16*8, 2*8*12 and 2*8*12 k-point meshes for the same three structures, respectively. In systems with occupied Nb 4$d$ orbitals (i.e., $SrNbO_3$ and $SrNbO_{3.4}$), we employed ferromagnetic ordering of the local moments on the $Nb^{4+}$ ions.

**Acknowledgements**

The authors appreciate Dr. Kokkoris from National University of Athens, Greece, for the carbon ions RBS measurement. Financial support from the Singapore-Berkeley Research Initiative for Sustainable Energy (SinBerRISE) program is gratefully acknowledged. The NUS researchers acknowledge Singapore National Research Foundation under its Competitive Research Funding "Control of exotic quantum phenomena at strategic interfaces and surfaces for novel functionality by in-situ synchrotron radiation" (NRF-CRP 8-2011-06), MOE-AcRF Tier-2 (MOE2015-T2-1-099), and FRC.

**Author contributions**

D. Y. W. and Y. L. Z. developed this project, designed the research strategy, and analyzed the results. Y. L. Z, D. Y. W. and B. X. Y. prepared the samples and measured the electronic transport properties. D. Y. W. carried out the transient absorption spectra measurements with the J. Q. C.'s instruction. Y. C. performed the DFT calculations. T. C. A. carried out ellipsometry spectroscopy characterization. Z. H. measured the RSMs. J. H. carried out the photocatalytic experiments. C. T. N. performed the TEM imaging. M. R. M. measured the RBS spectra. R. X. supervised the photocatalytic measurement. H. Z. analyzed the TEM images. A. supervised the RSMs experiments. A. R. supervised the ellipsometry spectra measurements. A. M. supervised the TEM imaging of the films. M. B. H. B. supervised the RBS experiments. M. A. supervised the DFT calculations. Q-H. X. supervised the transient absorption spectra measurements. D. Y. W and Y. L. Z. prepared this manuscript. T. V. supervised and led the project. All authors contributed to the scientific discussion and manuscript revisions.

**Competing financial interests statement**

The authors declare no competing financial interests.

**Additional information**

Supplementary information is available in the online version of the paper. Reprints and permissions information is available online at www.nature.com/reprints. Correspondence and requests for materials should be addressed to Y. L. Z. and T. V.

**Figure Captions:**

**Fig. 1 | XRD patterns and TEM image of SrNbO3 film grown on LaAlO3(001) substrate. The film was deposited under oxygen partial pressure of 5×10-6 Torr.** (a) θ-2θ scan, where * indicates the signals of the substrate. Inset shows the rocking curve. (b) RSMs on (-103), (103), (0-13) and (013) reflections from the substrate and the film. (c) High resolution TEM image of a square region. The shaped open circuit area by yellow arrows highlights an edge dislocation core (type a[100]p).

**Fig. 2 | The optical properties of $SrNbO_{3+x}$ films.** (a) UV-Visible-NIR spectra of $SrNbO_{3+x}$ thin films deposited at various oxygen partial pressures, an absorption edge located at the wavelength of 300nm can be observed. (b) The transmission, reflection spectra of the film deposited at $5\times10^{-6}$ Torr, with the absorbance spectrum obtained as plot in blue. Ellipsometry analysis of (c) The refraction index (n) and extinction coefficient (k) (d) Loss function and (e) Reflectivity of $SrNbO_3$ as a function of photon energy.

**Fig. 3 | Electronic transport properties and transient absorption spectra of $SrNbO_3$ films.** Temperature dependent transport properties of the films prepared at various oxygen pressures: (a) resistivity, (b) Mobile electron density obtained from Hall measurement, and (c) electron mobility of the films. The differential reflectance (ΔR/R) spectra for $SrNbO_3$ film of the delay time at delays of (d) 2.0 ps and (e) 1000.0 ps with pump light of various wavelengths with a white light continuum probe.

**Fig. 4 | The time-resolved transient absorption spectra of SrNbO3 thin films.** The excitation wavelength dependent time-resolved pump-probe dynamic spectra with the probe wavelength at 670 nm with the time range of 2.0 ps (a) and 1100.0 ps (b). The excitation wavelength dependent carrier lifetimes are shown for two processes with the probe pulse at 670 nm, (c) the fast process corresponding to the plasmon decay and (d) the slow process corresponding to the thermal dissipation process.

**Fig. 5 | Crystal structures and calculated band structures and projected density of states (DOS) of $SrNbO_3$, $SrNbO_{3.4}$ and $SrNbO_{3.5}$ using DFT+U (U=4eV).** (a) Distorted perovskite structure of $SrNbO_3$. $O^{2-}$ : small red sphere, $Sr^{2+}$: large green sphere, $Nb^{4+}$: small blue sphere. Unit cell is shown with black solid line. (b) Band structure and DOS of $SrNbO_3$, showing its metallic behavior. (c) Layered structure of $SrNbO_{3.4}$ with extra oxygen layers inserted every 5 octahedral layers. The dash lines indicate where the extra oxygen layers are. $Nb^{4+}$ (small blue sphere) and $Nb^{5+}$ (small green sphere) are given different colors to show charge ordering in this composition. (d) Band structure and DOS of $SrNbO_{3.4}$, with significantly reduced carriers at the Fermi level. (e) Layered structure of $SrNbO_{3.5}$, with an extra oxygen layer inserted every 4 octahedral layers. (f) Band structure and DOS of $SrNbO_{3.5}$, showing its insulating behavior.

**Figures:**

Fig.1

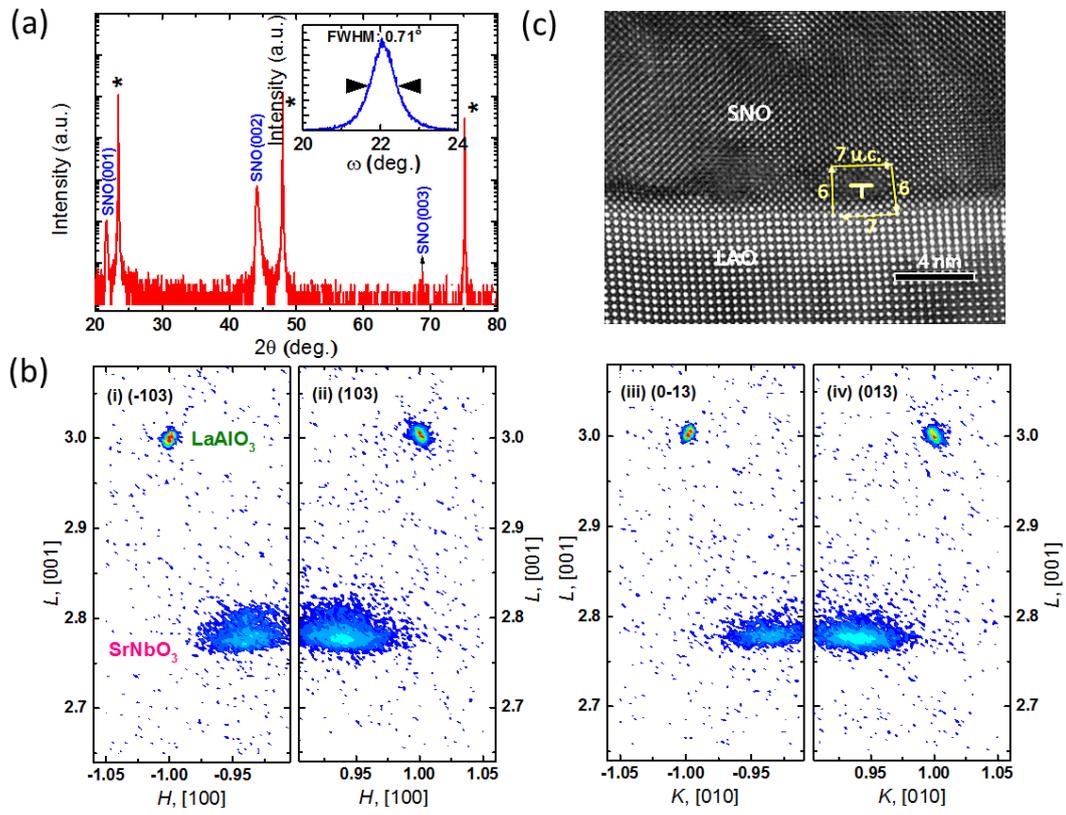

Fig.2

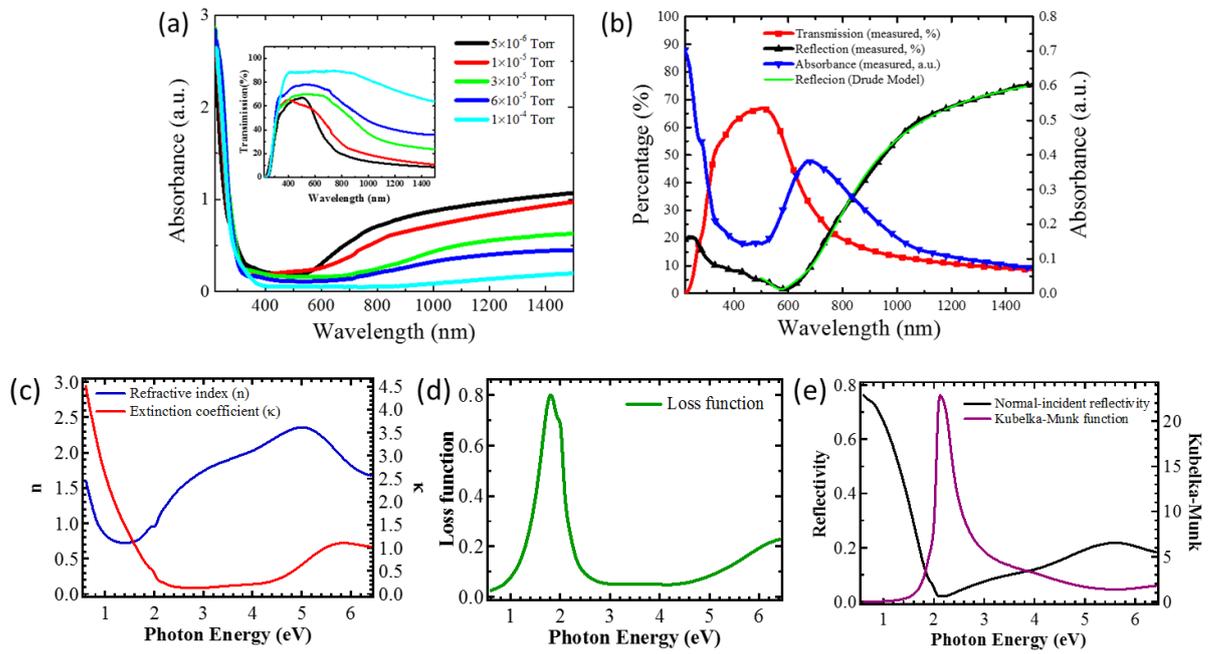

Fig.3

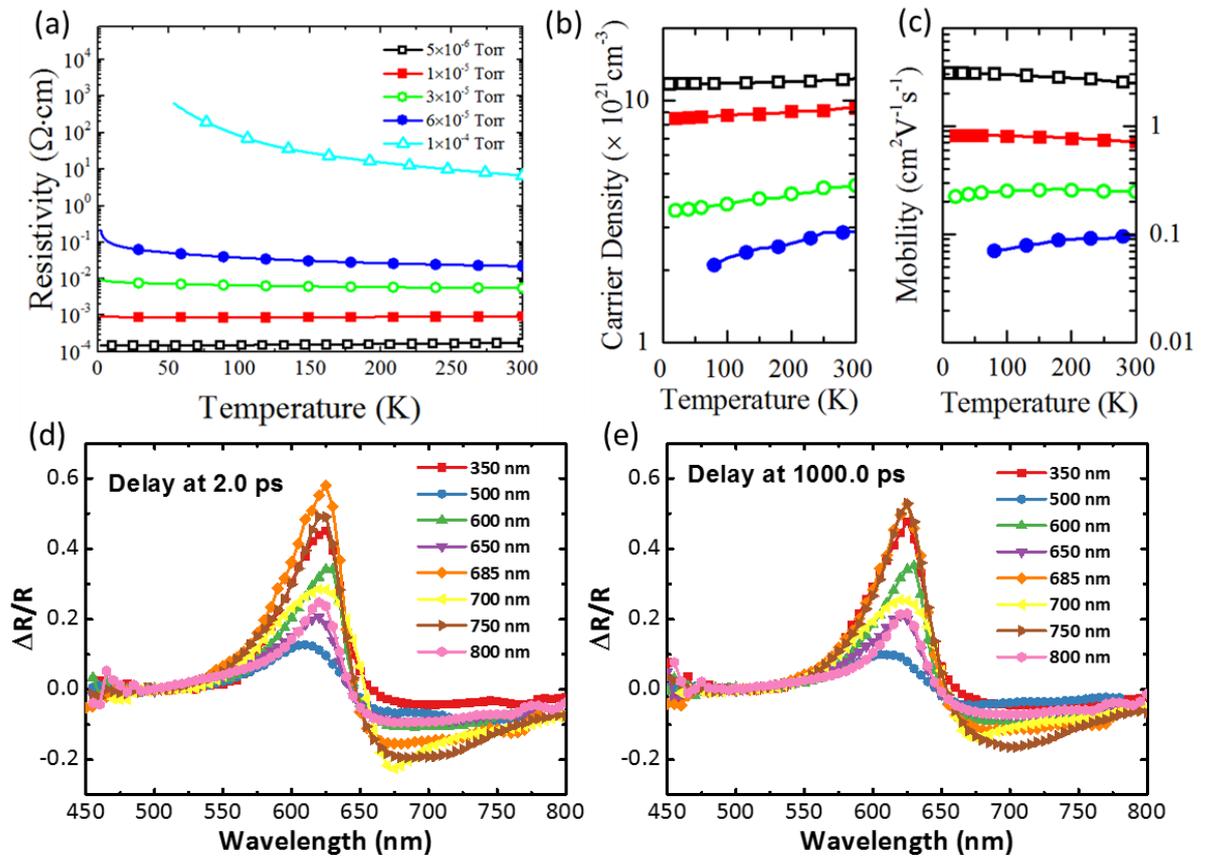

Fig. 4

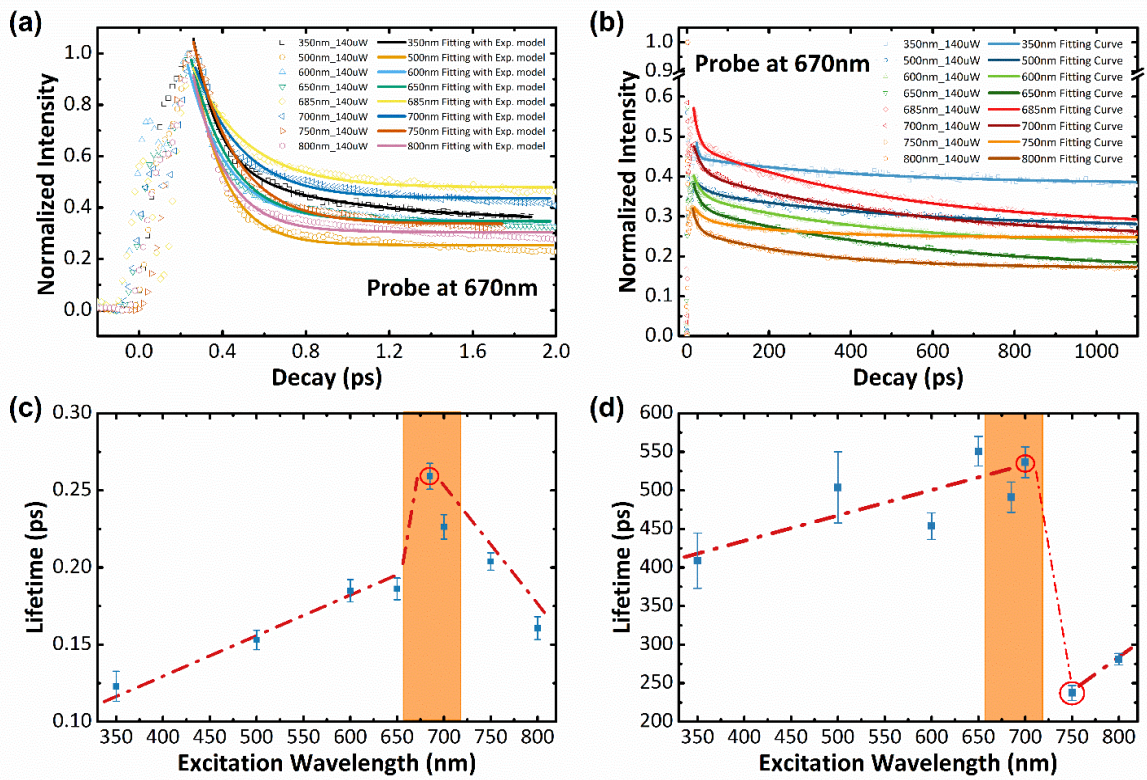

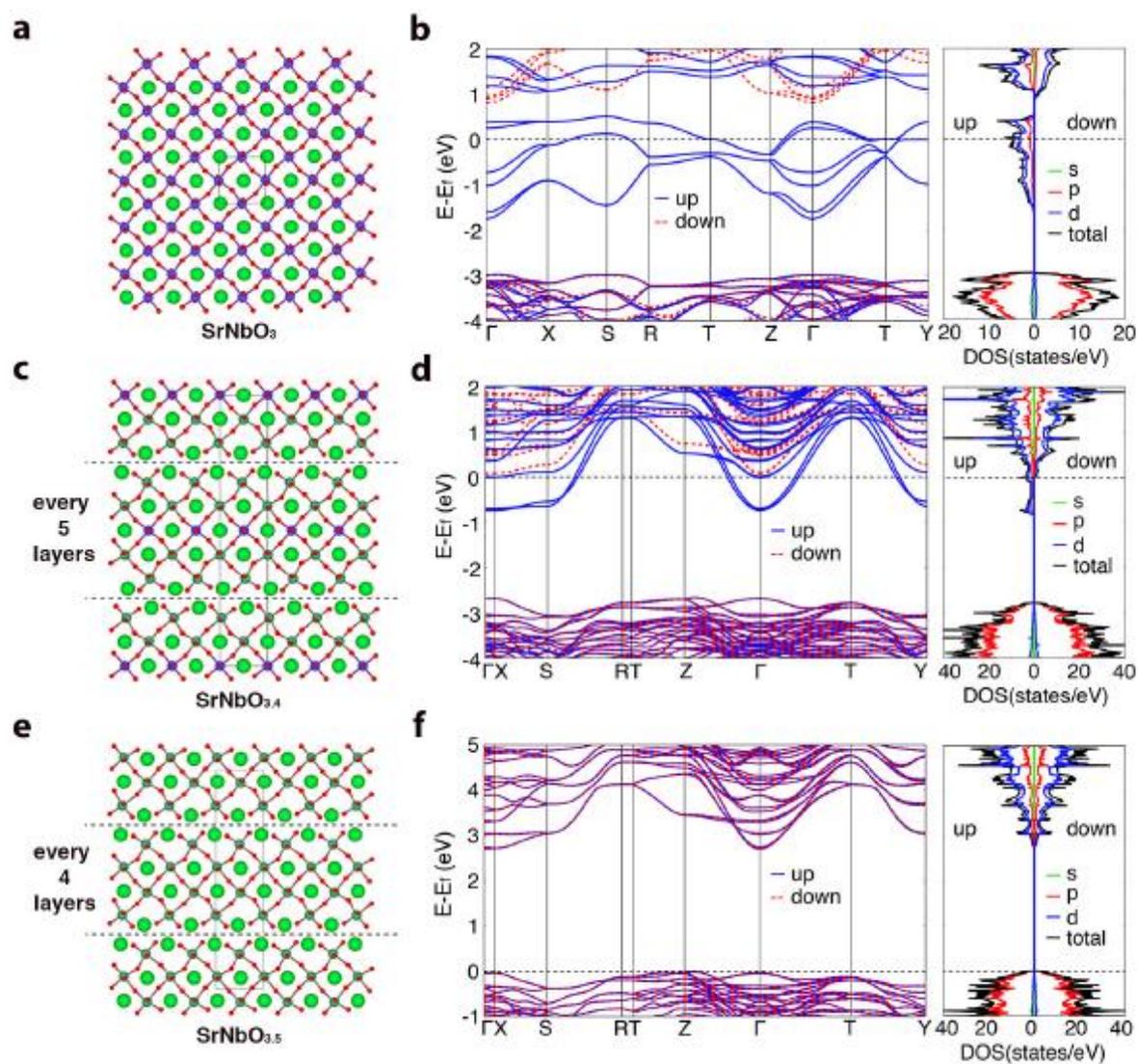